\documentclass[10pt,journal,comsoc]{IEEEtran}

\makeatletter
\def\endthebibliography{%
	\def\@noitemerr{\@latex@warning{Empty `thebibliography' environment}}%
	\endlist
}
\makeatother
\usepackage{amssymb}% http://ctan.org/pkg/amssymb
\usepackage{pifont}% http://ctan.org/pkg/pifont
\newcommand{\xmark}{\ding{53}}%
\usepackage[flushleft]{threeparttable}
\usepackage{gensymb}
\usepackage{color}
\usepackage{flushend}

\usepackage[pdftex]{graphicx}
\usepackage{tabularx,booktabs}
\usepackage{caption}
\usepackage {longtable}
\usepackage{multirow}
\usepackage{balance}
\newcolumntype{C}{>{\centering\arraybackslash}X} % centered version of "X" type
\usepackage{lipsum}

\usepackage{tikz}
\def\checkmark{\tikz\fill[scale=0.4](0,.35) -- (.25,0) -- (1,.7) -- (.25,.15) -- cycle;}
\ifCLASSOPTIONcompsoc
  \usepackage{cite}
\else
  \usepackage{cite}
\fi
\ifCLASSINFOpdf
\else
\fi
\usepackage{atbegshi}% http://ctan.org/pkg/atbegshi
\AtBeginDocument{\AtBeginShipoutNext{\AtBeginShipoutDiscard}}

\usepackage{cite}
\begin{document}

\title{A Survey on LoRa Networking: Research Problems, Current Solutions and Open Issues}

\author{Jothi~Prasanna~Shanmuga~Sundaram,~\IEEEmembership{Student member,~IEEE}, Wan~Du,~\IEEEmembership{Member,~IEEE}, Zhiwei~Zhao,~\IEEEmembership{Member,~IEEE}}

\thanks{ Wan Du and Jothi Prasanna Shanmuga Sundaram are with the Department of Computer Science and Engineering, the University of California, Merced. E-mail: \{wdu3, jshanmugasundaram\}@ucmerced.edu, Zhiwei Zhao is with the School of Computer Science and Engineering, University of Electronic Science and Technology of China. Email: zzw@uestc.edu.cn. Wan Du is the first corresponding author of this article and Zhiwei Zhao is the second corresponding author.}
\IEEEtitleabstractindextext{%
\begin{abstract}
\label{sec:abstract}
Wireless networks have been widely deployed for many Internet-of-Things (IoT) applications, like smart cities and precision agriculture. 
Low Power Wide Area Networking (LPWAN) is an emerging IoT networking paradigm to meet three key requirements of IoT applications, i.e., low cost, large scale deployment and high energy efficiency. 
Among all available LPWAN technologies, LoRa networking has attracted much attention from both academia and industry, since it specifies an open standard and allows us to build autonomous LPWAN networks without any third-party infrastructure. 
Many LoRa networks have been developed recently, e.g., managing solar plants in Carson City, Nevada, USA and power monitoring in Lyon and Grenoble, France. 
However, there are still many research challenges to develop practical LoRa networks, e.g., link coordination, resource allocation, reliable transmissions and security.
This article provides a comprehensive survey on LoRa networks, including the technical challenges of deploying LoRa networks and recent solutions.
Based on our detailed analysis of current solutions, some open issues of LoRa networking are discussed. The goal of this survey paper is to inspire more works on improving the performance of LoRa networks and enabling more practical deployments.
\end{abstract}

}

\maketitle
% \thispagestyle{empty}
%\pagestyle{plain}

\IEEEdisplaynontitleabstractindextext
\vskip 1cm
\IEEEraisesectionheading{\section{Introduction}
\label{SEC:INTRO}}

\IEEEPARstart{T}{he} Internet-of-Things (IoT) applications~\cite{lin2017survey,al2015internet}, like smart homes and smart cities, become more and more pervasive, which result in increasing density and scale of networked sensor deployments~\cite{stankovic2014research,Americas,Paper}.
Ericsson mobility report~\cite{Ericsson2017} states that connected IoT devices will grow from seven billion in 2017 to 20 billion in 2023, corresponding to an annual growth rate of 19$\%$. The IoT applications employ \textit{things} with sensing capabilities to sense the environment, communicate with other devices and humans and make intelligent decisions. To connect IoT devices, wireless networks are required to provide robust operations and wider coverage with high energy efficiency~\cite{lin2017survey}.
The IoT end devices are mostly battery powered. They are expected to work for a longer span of five to ten years without any maintenance.
These IoT end-devices are also expected to cover a large geographical area. For example, the forest monitoring application installs end devices throughout the forest region. 
The devices communicate small payloads to convey interesting data like humidity, temperature and other variables over a longer distance in a multi-hop manner. 

The above requirements have led to a new branch of IoT networking technology, called Low Power Wide Area Networking (LPWAN), as conventional IoT networking technologies like Zigbee and Bluetooth can only provide  a shorter range~\cite{gazis2017survey,al2015internet,chen2017machine}. LPWAN employs simple network topology and long distance communication with low data rates to attain high energy efficiency~\cite{raza2017low}. Existing LPWAN technologies can be divided into three categories i.e., networks based on cellular infrastructure \cite{wang2017survey,flore20163gpp}, networks using third-party infrastructure \cite{zuniga2016sigfox}, autonomous LPWAN networks without any third-party infrastructure \cite{vangelista2015long}.

First, existing cellular technology covers a wide area but its energy efficiency does not match LPWAN requirements as they were not commissioned for machine-type communications. As cellular networks are already densely populated, a new massive wave of IoT devices cannot be handled as it leads to heavy interference. To overcome these challenges, intensive research is being conducted on Cellular-IoT technologies like LTE-M~\cite{lee2016lte},~\cite{mezzavilla2018end}, NB-IoT~\cite{wang2017survey,flore20163gpp} and EC-GSM~\cite{flore20163gpp}.
For example, NB-IoT~\cite{wang2017survey,flore20163gpp} operates at licensed Long-Term Evolution (LTE) bands using Single-Carrier Frequency Division Multiple Access (SC-FDMA) for uplink and Orthogonal Frequency Division Multiple Access (OFDMA) for downlink modulation. It facilitates higher Quality-of-Service (QoS) \cite{sinha2017survey}.

Second, some service providers, like SigFox~\cite{zuniga2016sigfox}, Ingenu~\cite{ingenu2016works} and Weightless~\cite{webb2012understanding}, are proprietary networks. 
Ingenu~\cite{ingenu2016works} is a founding member of IEEE 802.15.4K task group. It leans on completing its stack whereas SigFox and LoRa focus on faster time to market. It operates at the 2.4GHz band. Ingenu uses Random Phase Multiple Access (RPMA) modulation which gives higher link budget and coverage while energy efficiency becomes a downside. Ingenu also suffers from interference of other technologies like WiFi, low structural penetration of signals and increased propagation loss at high frequencies~\cite{Alliance2015}.
SigFox~\cite{zuniga2016sigfox} is more popular in European region because of the traction made by widely available vendors like Axom, Texas Instruments and Silicon labs. 100Hz bandwidth (BW) and Ultra-Narrow Band technology are utilized for transmitting smaller packets (12 bytes, up to 140 messages per day) at low data rates (up to 100 bits per second) modulated with Binary Phase Shift Keying (BPSK). Major limitations of SigFox includes (i) being proprietary closed source technology, (ii) Low security mechanisms and (iii) restrictions on downlink transmissions \cite{fujdiak2018track}.

\begin{table*} [h]
	\caption{Comparison of LPWAN technologies}
    \label{TAB:COM}
    \small
	\begin{tabularx}{\textwidth}{@{}l*{10}{C}c@{}}
		\toprule
			                              &LoRa								&Ingenu	          							& Sigfox		& NB-IoT	   \\  
		\toprule
		Third-party infrastructure   	  & Open source						&Closed source     							& Closed source & Open Source   \\
		\midrule
		Operating Band   	              & ISM Sub-GHz						&ISM 2.4GHz        							& ISM Sub-GHz   & Licensed LTE band 180KHz           \\
		\midrule
		Channels					      &Multiple SF with 64+8 UL and 8 DL&40 1-MHz channels, 1200 signals per second	&360 channels   & 3 DL and 2 UL\\
		\midrule
		Modulation                        & CSS, FSK						& RPMA-DSSS, CDMA  							& DBPSK, GFSK   & OFDMA, SC-FDMA \\
		\midrule
		Data rate			     	      & 0.3-37.5 Kbps					&78 Kbps UL, 19.5 Kbps DL 					& 100 bps UL, 600 bps DL& up to 250kbps   \\
		\midrule
		Communication Range 	          & 5Km -15Km \cite{raza2017low,wang2017survey}& 15Km \cite{raza2017low,wang2017survey}&1-Km to 5-Km \cite{raza2017low,wang2017survey} & up to 35Km\cite{raza2017low,wang2017survey}\\
		\midrule
		Payload Length 				      & up to 250Bytes					&10 Kilobytes          						&12Bytes UL and 8B DL &1600 Bytes \\
		\midrule
		Authentication   		          & Symmetrical Authentication key  &Mutual Authentication						&Burnt-in symmetrical authentication key   &Mutual Authentication \\
		\midrule
		Encryption				          &AES 128bit    					&AES 256bit           						&\xmark & LTE encryption                       \\
		\bottomrule
	\end{tabularx}
\end{table*}

Finally, LoRa networking~\cite{bor2016lora} is widely used for LPWAN applications because, \textit{LoRa networking is an open-source technology that enables autonomous network set-up at low cost.} 
LoRa networks have been widely deployed for many applications and research systems. 
The openness of LoRa makes it an excellent choice for diverse IoT deployments~\cite{vangelista2015long}. General IoT applications include smart buildings~\cite{Semtech2017}, smart cities~\cite{Semtech2017a}, smart agriculture \cite{Semtech2017c}, smart meters~\cite{Semtech2017b},~\cite{magrin2017performance} and water quality measurement~\cite{du2014optimal,du2015sensor,du2017pando,du2015pipelines}.  

Major LPWAN technologies are compared in Table \ref{TAB:COM}. Acronyms used in this table are described in Table \ref{TAB:acronym}. 
Working in sub-GHz band using CSS modulation makes LoRa technology immune to interference as the chirp signal varies its frequency linearly with time. The chirp signals utilize the available bandwidth instantaneously consuming low power than the other LPWAN technologies. A nominal coverage of 5Km-15Km~\cite{wang2017survey} is obtained with higher payload (up to 250Bytes) when compared to other technologies. 
LoRa networks offer better downlink capabilities than Sigfox and Ingenu. LoRa networking provides light-weight encryption and authentication mechanisms that can be configured during activation. Another important advantage of LoRa networks is that the configuration and firmware updates can be sent over the air \cite{FUOTA}. 
% LoRa technology also provides a unique mechanism called Time-of-Arrival to localize the end-devices to a specific gateway \cite{bissett2018analysing}.

\begin{table}[t]
	\centering
	\caption{Acronyms found in this paper}
	\label{TAB:acronym}
	    \small
	\begin{tabular}{ l  l }
		\hline
		Acronym & Description\\
		\hline
		bps     & bits per second\\
		CDMA    & Code Division Multiple Access\\
		CR		& Code Rate\\
		CSS     & chirps Spread Spectrum\\
		DBPSK   & Differential Binary Phase Shift Keying\\
		DL      & Downlink\\
		DSSS    & Direct Sequence Spread Spectrum\\
		EC-GSM  & Extended Coverage-GSM\\
		FSK     & Frequency Shift Keying\\
		GFSK    & Gaussian Frequency Shift Keying\\
		Kbps    & Kilo bits per second\\
		LPWAN   & Low Power Wide Area Networks\\
		NB-IoT  & Narrow Band-IoT\\
		OFDM    & Orthogonal Frequency Division Multiple Access\\
		RPMA    & Random Phase Multiple Access\\
		SC-FDMA & Single Carrier-Frequency Division Multiple Access\\
		SF      & Spreading Factor\\
		UL      & Uplink \\		
		\hline
	\end{tabular}
\end{table}

\textbf{Why a new survey on LoRa networking?} Raza et.~al ascertain the need of LPWAN by justifying the inability of legacy wireless systems to comply with the constraints of LPWAN \cite{raza2017low}. The design goals of LPWAN along with various techniques to achieve these goals are discussed. On discussing the challenges and research directions, the authors find that most of the working groups focus on PHY and MAC layers.
We argue that the upper layers should also be discussed such as the efficient deployment of LPWAN. A brief description is provided on technical specifications of all LPWAN technologies while recent performance measurements, research challenges and solutions are not discussed in detail.

Sinha et~al. study two leading LPWAN technologies, LoRa and NB-IoT, by comparing their physical features, MAC protocol, QoS, latency, communication range and deployment cost of each technology \cite{sinha2017survey}. LPWAN application scenarios are categorized and some important parameters to be considered for each specific scenario are studied. Research challenges and recent technical advancements of each technology are not discussed in detail.

Different from the above mentioned surveys \cite{sinha2017survey, raza2017low}, our survey is focused on LoRa networks.
We study the recent performance measurements of LoRa networking \cite{fuidiak2018simulated, petajajarvi2015coverage, petajajarvi2017evaluation,harris2018development, Haxhibeqiri2017, mikhaylov2016analysis, magrin2017performance, Semspace0mmtech, bankov2016limits, Haxhibeqiri2017, angrisani2017lora, blenn2017lorawan, lauridsen2017interference, vejlgaard2017coverage, zhu2018evaluation, orfanidis2017investigating, ferre2017collision, aras2017exploring, butun2018analysis, miller2016lora, oniga2017analysis} to understand and devise a taxonomy for the research problems of LoRa networking. The recent solutions \cite{eletreby2017empowering, dongare2018charm, peng2018plora, talla2017lora, hessar2018netscatter, marcelis2017dare, blenn2017lorawan, bor2016lora, georgiou2017low, donmez2018security, sandell2017application, haxhibeqiri2017lora, reynders2018improving, abdelfadeel2018fair, reynders2017power, pop2017does, cuomo2017explora, van2017scalability, mikhaylov2017lorawan, cattani2017experimental, voigt2017mitigating, lee2017risk, kim2017simple, na2017scenario, P.Girard2015, kim2017dual, naoui2016enhancing, tomasin2017security} are further discussed in detail to understand the advancements of LoRa technology.
Finally, we present some open issues that could further improve the performance of LoRa networking.

A survey on LoRa technology has been recently published~\cite{haxhibeqiri2018survey}. It discussed the literature, solutions and open issues without any classifications. In our article, a clear taxonomy has been devised. Based on this taxonomy, the challenges, current solutions and open issues are discussed with tabulated version of system analysis and hardware experiments. The taxonomy provided in this article facilitates a clear understanding of the challenges, solutions and open issues. In addition, some of the recent contributions to LoRa networking \cite{eletreby2017empowering, dongare2018charm, demetri2019automated, el2019lorawan} are also discussed in our article. 

The rest of this paper is structured as follows. Section~\ref{SEC:TB} gives a brief description of LoRa technology. Section \ref{SEC:ED} lists the existing deployments of LoRa networks and their advantages. Section \ref{SEC:CHALLENGE} investigates the research challenges of LoRa networking and devises a taxonomy. Section \ref{SEC:SOLUTION} gives a comprehensive study on how these research problems are tackled by recent solutions. Section \ref{SEC:DIRECTION} discusses some open issues that still needs to be addressed and Section \ref{SEC:CONCLUSION} concludes the article.

\section{A Brief technical background of LoRa}
\label{SEC:TB}

\begin{figure}[t]
	\centering
	\captionsetup{justification=centering}
	\includegraphics[width= 0.49\textwidth, height=0.4 \textwidth, keepaspectratio]{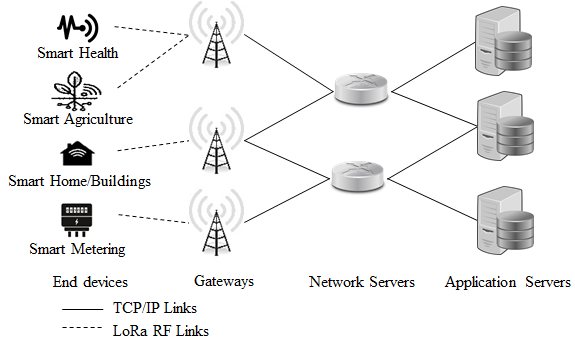}
	\caption{LPWAN Network architecture}
	\label{fig:net}
\end{figure}

\begin{figure}[t]
	\label{pic:RW}
	\centering
	\captionsetup{justification=centering}
	\includegraphics[width=0.5\textwidth, height=0.5\columnwidth, keepaspectratio]{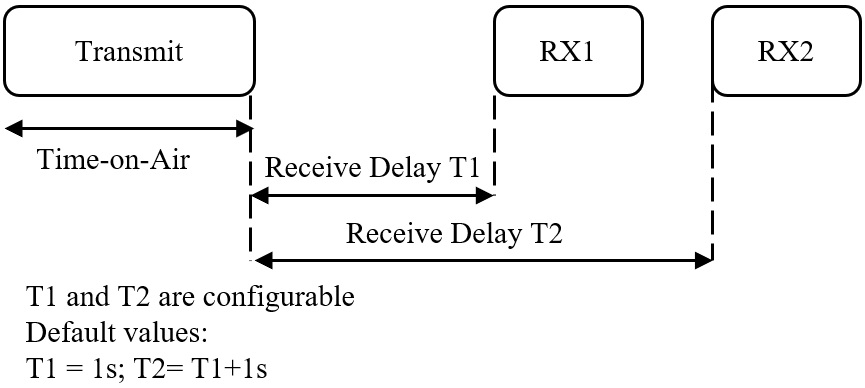}
%	\vspace{0.9 cm}
	\caption{Class A receive windows}
	\label{fig:carxw}
\end{figure}

\indent This section briefly describes the technical features of LoRa. LoRa operates in unlicensed sub-GHz ISM band (900MHz in USA and 860MHz in Europe). Using 125KHz, 250KHz and 500KHz of bandwidth, smaller payloads of up to 250 bytes can be transmitted over a distance of 5-15 Km and the system can last up to 5-10 years consuming low power according to the recent report \cite{wang2017survey}. A LoRa system comprises of end-devices, gateways, network and application servers. Figure \ref{fig:net} depicts the architecture of a typical LoRa system. end-devices collect information and send them to Gateways. Gateways relay messages between end-devices and network servers. A network server is configured to direct messages to appropriate application servers for processing. 
% All decisions are taken in the Application and network servers while gateway acts only as a relaying device. 
\indent 
There are three operating modes for LoRa. 
LoRa end-devices must implement Class A operating mode. Other optional modes like Class B and Class C can also be utilized. The end-devices operating in Class A and Class B modes are generally battery powered while the end-devices operating in Class C is mains powered. Class A utilizes less energy than Class B and C. In Class A, after sending confirmed messages, end-devices expect an acknowledgment (ACK) from the Network server during two pre-agreed time-slots known as ``receive windows (RW)''. Figure \ref{fig:carxw} depicts the RWs of Class A operating mode. Frequency and data rate of the first RW is the same as the uplink transmission parameters whereas the second slot operates on pre-agreed parameters to improve the robustness of message transmissions. end-devices do not expect replies from the server for unconfirmed messages. Class B operating mode opens additional receive windows scheduled by gateways through beacon packets. Class C mode has no downlink restrictions and can receive downlink messages any time whenever it is not in a transmitting state.  

In general, LoRa denotes the physical layer while LoRaWAN denotes the MAC layer communications and networking in LoRa stack.

\indent \textbf{LoRa.} The physical layer of LoRa technology uses Chirp Spread Spectrum (CSS). Chirps are the signals whose frequency varies linearly with time within the available bandwidth. This attribute makes the chirp signals resilient to noise, fading and interference. Every LoRaWAN packet starts with a preamble of ten chirps and six synchronization chirps followed by the data. Each chirp can modulate multiple chips (data bits). The number of data bits modulated depends on the parameter Spreading Factor (SF). For example, nine bits can be encoded in a chirp using SF9. A message sent with higher SF takes more time on air and reduces the data rate but improves resilience to noise. LoRa modulation also has two other parameters namely Bandwidth and Code Rate (CR). The bandwidth can be set to 125KHz, 250KHz and 500KHz and the CR can be set to 4/5, 4/6, 4/7 or 4/8. 

\indent \textbf{LoRaWAN.} The LoRa community often refers LoRaWAN as a ``MAC in the cloud'' design \cite{dongare2018charm}. Gateways are the forwarders acting based on commands from the servers. All MAC decisions like data-rate, handling ACKs are decided at the servers. LoRaWAN MAC employs two modes to divide air-time between end-devices for handling collisions. The first mode is the ALOHA MAC that allows end-devices to transmit as soon as they wakeup and exponential back-off is applied in case of collisions. The second mode is the TDMA scheduler where the network server allots time-slot for each end-device to transmit their messages.        

\subsection{Unique properties of LoRa}
LoRa technology has some unique properties making it a widely used technology. The unique properties are (i) Ultra-long distance, (ii) Low cost and complexity devices (iii) Long lifetime of nodes, (iv) concurrent reception capacity of gateways and (v) robustness in Doppler effect. All these unique properties are experimentally verified by \cite{liando2019known}. 

\textbf{Ultra-Long distance:} In Line-Of-Sight (LoS) communications, the longest SF12 can achieve a distance of up to 9Km with Packet Reception Ratio (PRR) $>$ 70\% and the smallest SF7 can achieve a distance of 5Km for PRR $>$ 70\%, according to the report in \cite{haxhibeqiri2017lora, petajajarvi2017evaluation}. In Non-Line-of-Sight (NLoS) scenarios comprising of buildings, the longest distance achieved is around 2Km \cite{el2019lorawan}. It is also noticable that the communication distance is affected by the parameters Bandwidth, SF, transmission power and coding rate \cite{angrisani2017lora}.

\textbf{Low cost and complexity:} The LoRa devices are fabricated such that they are not complicated hence reducing the price. Reduced complexity also reduces the overheads incurred during communications. For example, a sophisticated CSMA is not employed instead a CAD is employed that will just check for preambles in the channel before transmission. There is no signalling overhead like other traditional communication networks. Whenever a node wants to transmit, it wakes up, checks for channel status, transmits and goes back to sleep. 
% No complex mechanism has been employed.

\textbf{Long lifetime:} The LoRa consumes around 120-150 mW during transmission and 10-15 mW for MCU operations based on different radios and host-boards used. This can be extrapolated to 2-5 years in total life time while the duty cycle is varied from 0.1\% to 10\% \cite{liando2019known}. 

\textbf{Concurrent reception of gateways:} Current LoRa gateways are capable of concurrent reception on 8 channels. Even the same SF can be received on different channels. All the different spreading factors from SF7 - SF12 are orthogonal and transmissions with different SFs can be received on the same channel concurrently. 

\textbf{Robustness to Doppler effect:} Liando et al. \cite{liando2019known}  prove that LoRa signals are robust to Doppler effect. The CSS modulation used by LoRa is highly resistant to Doppler effect. Mobile LoRa end-devices at a constant speed and in LoS can yield PRR $>$ 85\% \cite{liando2019known}.

\section{Existing deployments of LoRa networks}
\label{SEC:ED}
\indent This section explains the existing deployments of LoRa networking and their advantages. There are many use cases like building management system \cite{Semtech2017}, smart agriculture \cite{Semtech2017c}, smart parking \cite{Semtspace0mmech} and smart lighting \cite{Semtech}. some popularly known real-world deployments are summarized in Table III. An overview of these deployments is discussed below.

\textbf{Smart cities and Urban Deployments:} Influx of population toward cities demands a better way of governing and organizing amenities for optimal usage. Semtech's white paper \cite{Semtech2017a} explains how LoRa LPWAN could provide efficient usage and governance to make cities smart. Some applications that could improve daily life of the people are Smart parking \cite{Semtspace0mmech}, Smart lighting \cite{Semtech}. These applications will improve people's living experience. Some in-field deployments \cite{Semtech2017a} are (i) waste management in Seoul, North Korea, (ii) integrated sensing of Solar power plants in Carson city, Nevada, USA, (iii) power monitoring in Lyon and Grenoble, France.

\begin{table}[t]
	\centering
	\label{TAB:IFD}
	\caption{In-field deployments of LoRa networks}
	\small
	\begin{tabular}{ c  c }
		\hline
		LoRa Deployment & Location \\
		\hline
		Waste  Management \cite{Semtech2017a} & Seoul, North Korea \\
		Solar power plant management \cite{Semtech2017a}& Nevada, USA \\
		Power usage monitoring \cite{Semtech2017a}& Lyon, France \\
		Power usage monitoring \cite{Semtech2017a}& Grenoble, France \\
		Smart meters \cite{Semtech2017b}& Gehrden, Germany \\
		Smart golf course \cite{Semtech2018} & Calgary, Canada\\
		Smart Islands \cite{Semtech2018a} & Mallaorca, Spain\\
		\hline
	\end{tabular}
\end{table}

\indent Seoul experiences humongous floating crowds every day. As the crowd moves through the city dynamically, probing the capacity of waste-bins became tedious. City management installed LoRa-enabled smart bins to periodically collect the capacity of waste-bins. This helped to clear the bins as soon as they are filled. This application gives a 66\% reduction in waste collection frequency, an 83\% reduction in costs and a 46\% increase in recycling. 

\indent Carson City management in Nevada found that effective transition between legacy and  solar power is important when solar efficiency reduces during cloudy climates and nights. LoRa-enabled monitoring system monitors the current environment status of the solar-deployed site. Decision to use solar or legacy power is based on this collected data. This system reduced 15\% of operational expenses and boosted solar power output to 75,000 kWh of clean power because of proper transition between solar and legacy power.

\indent In Lyon and Grenoble, power consumption monitors were deployed in households. This helped the residents to monitor their power usage and turn-off unwanted devices. This system helped to reduce power consumption by 16\%. 

\textbf{Smart meters:} Semtech's white paper \cite{Semtech2017b} describes and evaluates the capacity of LoRa technology for smart metering applications. This system is deployed in Gehrden, Germany where the population is 15,000. Around 7000 households were installed with LoRa-enabled smart meters. This application helped to reduce the man-power utilized for monitoring power usage by transmitting meter-readings periodically to the gateway. 

\textbf{Smart golf course:} Shaganappi point is a popular golf course in the city of Calgary, Canada serving a community of million people. Golfers play an average of 90,000 - 100,000 rounds of golf during April - November every year. It is identified that slow play devalues the overall experience. Improving this experience will help to retain customers. Hence, each golf cart is fixed with a LoRaWAN sensor. With real time movement and location of golf carts, pause of play anomalies are detected and appropriate help is provided to speed up the play. Large coverage of golf course requiring periodic updates with low power consumption makes LoRa a perfect solution for this use case. Hence overall experience of customer is elevated with maximization of revenue \cite{Semtech2018}.

\textbf{Smart Islands:} Mallorca is largest of Spain's five Balearic Islands popular for white sand and turquoise water. Currently there are 25 people for every meter of beach on the Island with 32\% anticipated growth by 2030. Citizens have shifted their attitude to conserve natural resources in the Island. LoRaWAN sensors are installed to aid water management systems. This periodically reports water quality and levels \cite{Semtech2018a}. This system has seen 25\% water savings since installation.

\begin{figure*}[t]
	\centering
	\captionsetup{justification=centering}
	\includegraphics[width=1\textwidth, keepaspectratio]{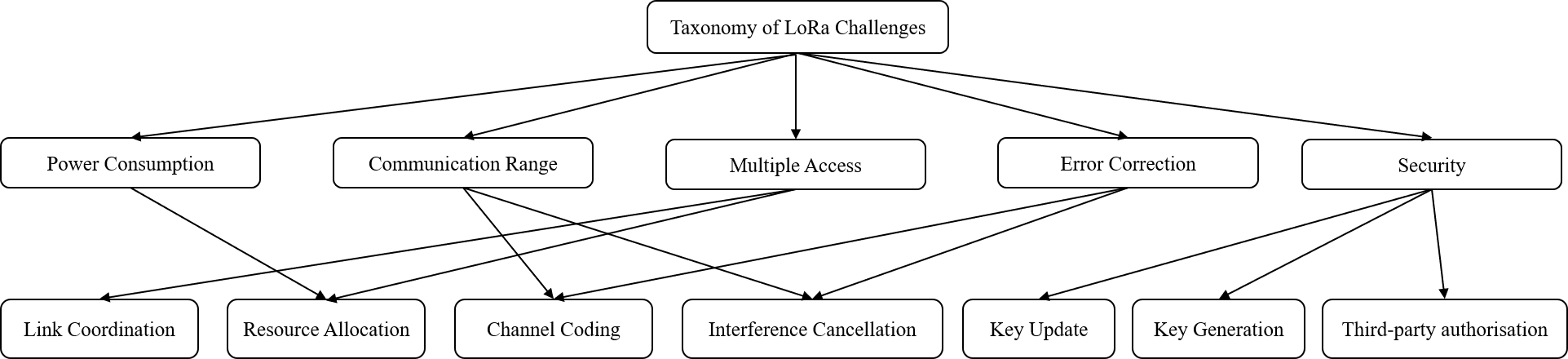}
	\caption{Taxonomy of LoRa Research Challenges}
	\label{pic:taxonomy}
\end{figure*}
\section{Taxonomy of research problems}
\label{SEC:CHALLENGE}

\indent In this section, we study the research challenges of LoRa networking. The severity of these challenges is identified by investigating its effect on the operations of LoRa technology. Finally, a taxonomy is devised to categorize these challenges. Figure 3 illustrates the taxonomy of the challenges of LoRa networking.

\subsection{Energy Consumption}

\indent The most important characteristic of LPWAN is its high energy efficiency. This becomes an important parameter to improve the longevity of end-devices. LoRa networks are expected to work for a longer period of 5-10 years with minimal maintenance. Hence, power consumption becomes a major challenge for LoRa networking. end-device operations can be classified into (i) micro-controller operations and (ii) wireless transmissions. Power consumed for micro-controller operations vary according to the chosen host board but the power consumed for wireless transmissions completely depends on the LoRa technology. Charm \cite{dongare2018charm} shows that wireless transmission extorts more power than micro-controller operations. LoRa technology employs two techniques to reduce energy consumption, (i) consuming instantaneous bandwidth for transmitting a chirp signal and (ii) not employing heavy MAC protocols for scheduling. In spite of these techniques, end-devices consume more power than expected due to some unavoidable circumstances like retransmissions caused by channel impairments. 

\subsection{Communication Range} \indent Large communication range is also an important rudiment of LoRa technology. Current LoRa technology relies on chirps spread spectrum, which is more resilient to interference. LoRa networking will be deployed in many scenarios such as homes, hospitals, schools, forest, etc. end-devices will be placed in the locations open to air, closed by concrete or steel, etc. Aiming such diverse deployment conditions, signal attenuation, propagation losses and fading have to be countered to improve signal penetration thus improving the coverage of LoRa networks \cite{harris2018development}, \cite{fuidiak2018simulated}. It has been noted that gateways can detect signals below a given threshold but cannot decode them. Devising a technique to decode these signals will improve communication range. Another important challenge is estimating the coverage of LoRa networks. Chall et al. \cite{el2019lorawan} study different models through empirical measurements in Lebanon. Demetri et al. \cite{demetri2019automated} identified that LoRa's signal coverage is anisotropic. This is because LoRa signals travel a longer distance and experience varying environments with dynamic and static obstacles in different directions. Mathematical models and systems for link quality estimation are still unexplored for LoRa. 

\subsection{Multiple Access:} \indent LoRa networking aims to connect thousands of end-devices to the network, communicating over a confined region and spectrum. Possibilities for these end-devices transmitting data concurrently varies based on the application. Multiple access is to allow multiple end-devices to share the limited spectrum for communication. The multiple access problems involve two different aspects namely Link coordination and Resource allocation.

\textbf{Link coordination:} Deploying thousands of devices require multiple-access to improve concurrent transmissions and avoid collisions. Links are coordinated through MAC protocols. The LoRa networking employs ALOHA and TDMA scheduler to coordinate links. These techniques cannot handle collisions while thousands of devices are connected to the network \cite{lauridsen2017interference}, \cite{vejlgaard2017coverage}. Thus, new techniques for handling collisions and coordinating links are required. This will help to upscale the density of LoRa deployments~\cite{zhu2018evaluation},~\cite{orfanidis2017investigating}.

\textbf{Resource Allocation:} In LoRa technology, the transmission is controlled by the parameters Spreading Factor (SF), Transmission power (TP), Bandwidth (BW) and Channel. Varying these parameters result in different transmission qualities. This can be leveraged to improve concurrent transmissions. Dynamically allocating reasonable resources to end-devices based on the deployed environment improves multiple access and thus scalability \cite{qin2017resource}. 

\subsection{Error Correction} \indent LoRa technology communicates data over long distances. While the message is transmitted over the air, it is possible for the data to get corrupted or lost due to channel effects, environmental conditions or collisions. Existing error correction schemes of LoRa networking, like hamming code, cannot aid data corruption or loss efficiently \cite{marcelis2017dare}. LoRa technology offers different spreading factors to make the signal more resilient to noises. SF12 is the stronger spreading factor but it takes more time-on-air. Even these signals can also be corrupted in dense environments \cite{petajajarvi2015coverage}, \cite{petajajarvi2017evaluation}. There are two types of current solutions namely (i) channel coding and (ii) interference cancellation. 

\textbf{Channel coding}. A recent channel coding technique proposed is DaRe \cite{marcelis2017dare}. DaRe is an application layer coding to retrieve lost data using redundant data. Sandell et al. \cite{sandell2017application} explain that this technique cannot aid bursty packet loss and has some limitations that bounds the performance. 

\textbf{Interference Cancellation:} Even though channel coding aids error, they cannot guarantee error correction in the case of collided signals. Interference Cancellation will extend error correction by untangling and extracting data from collided signals. Recently proposed technique Choir \cite{eletreby2017empowering} and Netscatter \cite{hessar2018netscatter} cancels interference but Choir's \cite{eletreby2017empowering} limitation is pointed out by Netscatter \cite{hessar2018netscatter}. Netscatter \cite{hessar2018netscatter} can cancel collisions of 256 concurrent transmissions, which is not adequate to handle thousands of end-devices by a single gateway.
For example, a large-scale temperature monitoring system requires all end-devices to transmit data at the same time \cite{Semtech2017b}. Hence, Interference cancellation is still an open problem that should be addressed to improve the performance of LoRa Networking.

\subsection{Security} \indent For all computer communications, security is a major concern. There are many security attacks like eavesdropping, selective forwarding and node impersonation \cite{zhou2008securing}. All the above mentioned attacks try to obtain the key used for encryption. If this key is compromised, the entire system can be broken. Currently, LoRa technology uses a symmetric key cryptographic technique with AES-128 bit encryption. Existing LoRa technology generates the key and never updates it. Hence, \textit{Key generation} and \textit{Key update mechanism} is a major concern.  \textit{Third-party authorization} is required when the application and network service providers are different. So, these applications require third-party authorization to ensure privacy and 
security~\cite{naoui2016enhancing,tomasin2017security,aras2017exploring}.

\section{Current Solutions}
\label{SEC:SOLUTION}
\indent In this section, we summarize the recent measurements and current solutions proposed to address the problem power consumption, communication range, error correction, multiple access and security. 
% There are two types of current solutions for Multiple access and Error correction problems. The solutions for multiple access can be classified into (i) Link coordination and (ii) resource allocation. The solutions for error correction are classified as (i) channel coding, (ii) interference~cancellation.\\  

\indent Existing works can be classified into two categories as performance measurements and current solutions tackling the above challenges. All the recent experiments, measurements and simulations use Class A end-devices unless specified. Table \ref{TAB:PF} summarizes performance measurements and case studies conducted on LoRa. Table \ref{TAB:PF} differentiates the measurements made through theoretical analysis, test-bed and simulated evaluations. It is to be noted that no performance measurements have been made on power consumption and error correction capabilities. Table \ref{TAB:RS} enumerates recent solutions to improve the performance of LoRa networking. The current solutions addressing more than one problem can also be identified in Table \ref{TAB:RS}.   

\begin{table*}[ht]
  \begin{threeparttable} 	
	\caption{Performance measurements of LoRa}
	\label{TAB:PF}

	\begin{tabularx}{\textwidth}{@{}l*{20}{C}c@{}}
		\toprule
		Articles	& \multicolumn{3}{c}{Communication Range} & \multicolumn{3}{c}{Packet Delivery} & \multicolumn{3}{c}{Multiplexing} & \multicolumn{3}{c}{Security} \\  
														    &T/M       &S       &T                     &  T/M   &S    &T                  &  T/M  &S  &T                     &T/M  &S  &T   \\ 
		\toprule                           
		Harris et al.\cite{harris2018development}           &          &          &\checkmark            &				                    &				                   &			\\
		\midrule
		Fuidiak et al.\cite{fuidiak2018simulated}           &          &\checkmark&                      &                                  &              &              \\
		\midrule
		Petajajarvi et al.\cite{petajajarvi2015coverage}    &\checkmark&&\checkmark&\checkmark&&\checkmark   &              &                  \\ 
		\midrule
		Petajajarvi et al. \cite{petajajarvi2017evaluation} & &&\checkmark        &&&\checkmark       &              &                  \\
		\midrule
		Haxhibeqiri et al.\cite{Haxhibeqiri2017}            & &&\checkmark          &                 &              &                      \\
		\midrule
		Mikhaylov et al.\cite{mikhaylov2016analysis}        & \checkmark &&         &                 &\checkmark    &                   \\	
		\midrule
		Magrin et al. \cite{magrin2017performance}          &                     &&\checkmark&       &              &        \\
		\midrule
		Semtech White paper \cite{Semspace0mmtech}          &                     &&&\checkmark       &              &        \\ 
		\midrule
		Bankov et al.\cite{bankov2016limits}                &                     &&\checkmark&       &              &        \\
		\midrule
		Haxhibeqiri et al.\cite{Haxhibeqiri2017}            &                     &&&\checkmark       &              &        \\
		\midrule
		Angrisani et al.\cite{angrisani2017lora}            &                     &&&\checkmark       &              &        \\
		\midrule
		Blenn et al.\cite{blenn2017lorawan}                 &                     &&&\checkmark       &              &        \\ 
		\midrule
		Lauridsen et al.\cite{lauridsen2017interference}  & &&                    &  &&               &&&\checkmark    & \\
		\midrule
		Vejlgaard et al.\cite{vejlgaard2017coverage}      &   &&                  &    &&             &&\checkmark&    & \\
		\midrule
		Zhu et al.\cite{zhu2018evaluation}            &    &&                 &      &&           &&\checkmark&    & \\
		\midrule
		Orfanidis et al.\cite{orfanidis2017investigating} &   &&                  &        &&         &&&\checkmark    & \\
		\midrule
		Ferre et al.\cite{ferre2017collision}            &   &&                  &          &&       &\checkmark&\checkmark&    & \\
		\midrule
		Aras et al. \cite{aras2017exploring}                &   &&                  &           &&      &   &&           &\checkmark&& \\
		\midrule
		Butun et al.\cite{butun2018analysis}				&&&					   &      &&           &   &&           &\checkmark	&&	\\
		\midrule
		Miller et al.\cite{miller2016lora}				&&&					   &     &&            &  &&            &\checkmark	&&	\\
		\midrule
		Oniga et al.\cite{oniga2017analysis}              &	&&				   &      &&           & &&             &&&\checkmark		\\
		\midrule
		Liando et al.\cite{liando2019known}              &	&\checkmark&\checkmark				   &      &\checkmark&\checkmark           & &\checkmark&\checkmark             &&&		\\
		\bottomrule
	\end{tabularx}
      \begin{tablenotes}
      % \begin{flushleft}	
      	\small
      	\item Table notes
      	\item T/M - Theoretical / Mathematical Analysis
      	\item S - Simulated evaluation
      	\item T - Testbed evaluation
      % \end{flushleft}
      \end{tablenotes}
      
  \end{threeparttable}
\end{table*}

\begin{table*} [ht]
	\caption{Summary of recent solutions for LoRa Challenges}
	\label{TAB:RS}
	\begin{tabularx}{\textwidth}{@{}l*{10}{C}c@{}}
		\toprule
		Article                                      &Multiplexing&Power Consumption&Communication Range&Error Correction&Security\\  
		\midrule
		Choir \cite{eletreby2017empowering}&\checkmark  &                   &  \checkmark       &               &          \\
		\midrule
		Charm \cite{dongare2018charm}       &\checkmark  & \checkmark            &  \checkmark       &               &          \\
		\midrule
		LoRa Backscatter \cite{talla2017lora}            &            & \checkmark        &\checkmark         &               &          \\
		\midrule
		PLoRa \cite{peng2018plora}             &            & \checkmark        &\checkmark         &               &          \\
		\midrule
		Netscatter \cite{hessar2018netscatter}    &\checkmark  & \checkmark        &                   &               &          \\
		\midrule
		DaRe \cite{marcelis2017dare}      &            &                   &                   & \checkmark    &          \\
		\midrule
		Blenn et al. \cite{blenn2017lorawan}         &\checkmark  &                   &                   &               &          \\
		\midrule
		Bor et al. \cite{bor2016lora}                &\checkmark  &                   & \checkmark        &               &          \\
		\midrule
		Chall et al. \cite{el2019lorawan}                &            &                   & \checkmark        &               &\\
		\midrule
		Demetri et al. \cite{demetri2019automated}   &            &                   & \checkmark        &               &\\
		\midrule
		Georgiou et al. \cite{georgiou2017low}       &\checkmark  &                   &                   &               &          \\
		\midrule
		Donmez et al. \cite{donmez2018security}      &            &                   &                   &               &\checkmark\\
		\midrule
		Sandell et al. \cite{sandell2017application} &            &                   &                   & \checkmark    &          \\
		\midrule
		Haxhibeqiri et al. \cite{haxhibeqiri2017lora}&\checkmark  &                   &                   &               &          \\
		\midrule
		Reynders et al. \cite{reynders2018improving} &\checkmark  &                   &                   &               &          \\
		\midrule
		Abdeel et al. \cite{abdelfadeel2018fair}     & \checkmark &                   &                   &               &          \\
		\midrule
		Reynders et al. \cite{reynders2017power}     & \checkmark &                   &                   &               &          \\
		\midrule
		Pop et al. \cite{pop2017does}                &\checkmark  &                   &                   &               &          \\
		\midrule
		Cuomo et al. \cite{cuomo2017explora}         & \checkmark &                   &                   &               &          \\
		\midrule
		Van et al. \cite{van2017scalability}         &\checkmark  &                   &                   &               &          \\
		\midrule
		Mikhaylov et al. \cite{mikhaylov2017lorawan} &\checkmark  &                   &                   &               &          \\
		\midrule
		Cattani et al. \cite{cattani2017experimental}&\checkmark  &                   &                   &               &          \\
		\midrule
		Voigt et al. \cite{voigt2017mitigating}      &\checkmark  &                   &                   &               &          \\
		\midrule
		Lee et al. \cite{lee2017risk}                &            &                   &                   &               &\checkmark\\
		\midrule
		Kim et al. \cite{kim2017simple}              &            &                   &                   &               &\checkmark\\
		\midrule
		Na et al. \cite{na2017scenario}              &            &                   &                   &               &\checkmark\\
		\midrule
		Girard \cite{P.Girard2015}                   &            &                   &                   &               &\checkmark\\
		\midrule
		Kim et al. \cite{kim2017dual}                &            &                   &                   &               &\checkmark\\
		\midrule
		Naoui et al. \cite{naoui2016enhancing}       &            &                   &                   &               &\checkmark\\
		\midrule
		Liando et al.\cite{liando2019known}			 &\checkmark  &					  &					  &				  &			 \\
		\bottomrule
	\end{tabularx}
\end{table*}

\subsection{Energy Consumption}
\label{SS:PC}

\indent Various techniques are employed to improve battery lifetime of the end-devices. Some works propose techniques (i) to harvest ambient energy from the environment \cite{orfei2016vibrations}, \cite{talla2017lora}, \cite{peng2018plora}; (ii) to use backscatter signals for transmission, \cite{talla2017lora}, \cite{peng2018plora},  and (iii) to detect and decode weak signals and increase data rate to reduce power consumption \cite{dongare2018charm}.     

\indent LoRa Backscatter \cite{talla2017lora} proposes a backscatter system for LoRa based on CSS modulation. Data can be transmitted up to 2.8 Km while consuming only 9.25 $\mu$watts of power at the rate of 37.5 Kbps. Power consumption is reduced by nearly 1000$\times$ than standard LoRa technology. These passive RF chips can be powered through solar panels attached to them. This technique is also analyzed over home/office sensing, precision-sensing of agriculture devices and epidermal devices to prove their efficiency.

\indent PLoRa \cite{peng2018plora} proposes a hardware and software co-design to enable battery-free LoRa networks, operating on the energy harvested from solar devices. The proposed PLoRa tag transmits data by backscattering ambient LoRa transmissions without external excitation signals unlike LoRea \cite{varshney2017lorea} and LoRa Backscatter \cite{talla2017lora} which uses dedicated hardware for generating excitation signals. The active LoRa signal emitted by a gateway or a node is converted into Passive LoRa signals to send data using ON-OFF keying technique. The power consumption of PLoRa is 250$\times$ smaller than the standard LoRa technology.

\indent Charm \cite{dongare2018charm} improves the battery lifetime up to 4x the standard LoRa technology by avoiding retransmission of the weak signals. This technique is discussed in subsection \ref{SS:CR}. FADR \cite{abdelfadeel2018fair} reduces the power consumption of standard LoRaWAN by 22\%. Wireless power charging has been a promising solution to handle energy consumption problem for wireless sensor networks \cite{navarro2018integration}. Realizing them on low-cost LoRa hardware has been done in \cite{Ideetron}. A circuit has been designed to enable wireless power transfer on LoRa enabled sensor nodes.

\indent Gao et al. \cite{gao19} investigated energy fairness problems in LoRa networks. Due to large differences between data rates used by different end-devices, end-devices far away from the gateway have to use a low data rate and spend more energy to transmit certain amount of data. To make energy consumption more fair across all end-devices in a LoRa network, Gao et al. \cite{gao19} propose to deploy more gateways to allow end-devices to use high data rates to reach at least one gateway. To make the network transmission more efficient, a network model is developed and used to allocate network resource to each end-device. A heuristic search based network resource allocation algorithm is developed to find the best network setting for each end-devices.  

\textbf{Key Insights.} Ambient energy harvesting is used to make LoRa devices battery free \cite{peng2018plora}. One of the most power consuming operation in LoRa wireless transmissions is generating carrier signals \cite{dongare2018charm}. One way to reduce this part of power consumption is to utilize the backscatter signals \cite{talla2017lora}. Another method is to leverage passive chips \cite{hu2016braidio} for carrier generation. Charm \cite{dongare2018charm} identifies that LoRa is able to receive weak signals but not able to decode them. Wireless power transfer \cite{navarro2018integration} is not feasible owing to the complex, costly hardware extensions in low power, low cost LoRa modules. Placing more gateways and dynamic allocation of TP addresses energy consumption problem and improves network lifetime.

\subsection{Communication Range}
\label{SS:CR}
\textbf{Testbed Measurements.} Semtech's white paper \cite{Semspace0mmtech} evaluates the capacity of LoRaWAN in dense urban environments. Ten gateways operating on 8 channels were used for trials with 100 sensors as end-devices to transmit at different data rates to mimic like 10,000 end-devices. Three phases of experiments with varying quantity of packet transmission were conducted to evaluate packet delivery. The first phase contains 250,000 volume of packets being transmitted at the rate of 104 packets per hour. The second phase generates 500,000 packets at the rate of 209 packets per hour and the third phase generates 1,000,000 packets at the rate of 417 packets per hour. For all the three phases, the achieved delivery rate is more than 95\%. This paper notes that multiple gateways will scale-up the network and improve communication range as end-devices can communicate with more than one gateway.

\indent Navarro et al. \cite{navarro2018integration} and Haxhibeqiri et al. \cite{Haxhibeqiri2017} evaluated communication range of LoRa in Industrial environments. The industrial environment at Royal Flora Auction Center, Netherlands, covering $250000 m^2$ of both indoor and outdoor spaces. LORANK \cite{Ideetron} Gateway was fixed 6 m above the floor and WiMOD-IM880A \cite{WIMOD2017} end-devices were attached to the trolleys 1.7 m above the ground. The nodes were triggered to transmit at SF7 and SF12. Measurements were taken from 43 measuring points covering indoor, outdoor spaces. Fifty packets were sent from each test point. Measurements show that a maximum of 6000 end-devices can be handled by a single gateway with a packet loss rate less than 10\%. Packet loss is around 6\% when less than 3500 end-devices are used.

\indent Petajajarvi et al. \cite{petajajarvi2015coverage} analyse the range of LoRaWAN with 14 dBm Transmit Power (TP) and the largest SF at Oulu, Finland. The experiment is conducted for 14 days during spring and summer. The population of Oulu is around 200,000 people with high rise buildings. Throughout the experiment, Kerlink’s LoRa gateway \cite{Kerlink2016} is fixed on a tower 24 m above the sea level with -137 dBm sensitivity in order to find the maximum communication range. Semtech 1272 transceiver \cite{Semtech2013} is used as an end-device. For on-ground measurements, end-devices are fixed on a car's roof-rack, approximately 2 m above the ground, which drove around major cities at 40 Km/h - 100 Km/h. For on-water measurements, end-devices are fixed on a radio mast of the boat. end-devices send packets periodically including GPS coordinates. The SF of end-devices is set to SF12 because the goal was to find the highest possible coverage. Maximum range noted on ground and water is 15 Km and 30 Km respectively. Total packet loss ratio for on-ground measurements and on-water measurements become 34\% and 32\% respectively. With these measured data, channel attenuation model is calculated for areas similar to Oulu.

\textbf{Research solutions.} Du et al. \cite{du2016rateless, du2014rateless} proposed a solution to improve the communication range of sparse wireless sensor networks. Choir \cite{eletreby2017empowering} identifies an intrinsic property of LoRa radios in which the carrier frequency varies by a small bound (902.4 MHz instead of 902.7 MHz) because of cheap radios. This is exploited to disentangle collided signals and extend the range up to 2.64$\times$ the standard LoRa technology. The nodes that are far away from the gateway transmit signals whose SNR goes below the noise floor.  It is assumed that the neighbouring nodes send data that do not vary to a greater extent. These physically co-located nodes, far away from the gateway, are coarsely synchronized through Class B beacons to transmit data at the same time to enable constructive interference thus improving the SNR above the noise floor. Gateways can receive and decode this collided signal to obtain approximate data of the region far away from gateways. This technique improves the communication range by 2.64$\times$ the standard LoRa technology.  

\indent Charm \cite{dongare2018charm} proposes a new hardware and software co-design to extend coverage and battery life of LoRa devices. This is achieved by allowing multiple gateways to send weak signals (that cannot be decoded by a single gateway) to cloud and coherently combine them to decode data. Programmable auxiliary hardware attached to the gateway improves the gateway's ability to detect very weak signals which cannot be directly detected by gateways. Joint decoding algorithm uses a heuristic approach to select signals to be combined at the cloud. Results show that the range is improved 3$\times$ and battery life is improved 4$\times$ the standard LoRa technology.

\indent LoRa Backscatter \cite{talla2017lora} discussed in subsection \ref{SS:PC} can send data to a receiver located 2.8Km away. Sensor Networks Over Whitespaces (SNOW) \cite{saifullah2016snow} was first designed for sensor networks to be connected over a wide area. As the traditional sensor networks cannot communicate to a longer distance, TV whitepsaces that could communicate a long distance is exploited. Scalability and energy efficiency is achieved by splitting carrier into several sub-carriers with parallel packet receptions. The PHY layer handles OFDM modulation and the MAC layer handles sub-carrier allocation. This technique is extended by Rahman and Saifullah \cite{rahman2018integrating} to integrate multiple LPWANs and improve the communication range specifically in infrastructure-restricted rural areas. The nodes located far away communicate with gateways over white spaces meant for TV signal communication. This technique is implemented on generic GNU radios. Its function on LoRa devices is still ambiguous.

\textbf{Key Insights.} The above measurement experiments throw light on (i) placing more gateways to improve network density, coverage and reduce energy consumption (ii)  less packet loss in networks with sparse end-device placements \cite{petajajarvi2015coverage}. 
Besides, we can infer the following insights:
\begin{itemize}
\item Radio imperfections occur in LoRa due to cheap radios put into use. 
\item These radios generate slightly different carrier frequencies than specified \cite{eletreby2017empowering}. \item LoRa gateways can receive weak signals but cannot decode them \cite{dongare2018charm}. 
\item Extending coverage of LPWAN's is experimented with GNU radios but its operation on commercial LoRa devices has not yet been studied \cite{rahman2018integrating}.  
\end{itemize}   

%\begin{figure}[t]
%	\centering
%	\captionsetup{justification=centering}
%	\includegraphics[width=0.5\textwidth, height=0.5\textwidth, keepaspectratio]{fi}
%	\vspace{0.3 cm}
%	\caption{First issue described in \cite{bankov2016limits}}
%	\label{fig:fi}
%\end{figure}

\subsection{Error Correction} 
In this subsection, the research solutions for error correction are classified into channel coding and interference cancellation. 
The solutions are explained in detail before concluding with key insights.  

\subsubsection{Channel Coding}
% \textbf{Research solutions.}
A new application layer data recovery technique called DaRe \cite{marcelis2017dare} is proposed based on Convolutional and Fountain codes. This technique extends data with redundant information. These redundant data are chosen from the previous data units so that the lost frame can be calculated from the other received frames. The disadvantage is that previous data units should be buffered in the memory for computing redundant information. This makes the generator matrix banded thus inducing difficulty to create degree of distribution according to LT codes. Sandell et. al \cite{sandell2017application} show that the memory affects performance and complexity. While \textit{DaRe} uses a complex Gaussian elimination making the decoding process complex, an optimized decoding technique is proposed in \cite{sandell2017application}. 

\indent Sandell et al. \cite{sandell2017application} analyses the technique proposed in \cite{marcelis2017dare} and shows that for larger packet loss probabilities, reducing the code rate increases the interference leading to reduced efficiency of data recovery. Showing the relationship between latency of decoding algorithm and packet loss probability, a less complex decoding algorithm like Accumulative Gaussian Elimination is proposed by Du et. al \cite{du2014rateless} to reduce latency. Finally, the paper concludes that data recovery through redundancy techniques increase number of transmissions and thus collision. So it cannot be used solely to aid packet loss. 

\subsubsection{Interference Cancellation}
% \textbf{Research solutions.} 
Choir \cite{eletreby2017empowering} and FTrack \cite{SenSys19_FTrack}  propose novel solutions to disentangle and decode collided signals. Using constructive interference, a sparse overview of the data is obtained from a group of geographically co-located end-devices far away from the gateway. Test bed evaluation shows that throughput is improved 6.84$\times$ than the standard LoRa. It is also proved that \textit{Choir} yields better results than multiple antenna deployments.

\indent NetScatter \cite{hessar2018netscatter} improves interference cancellation. It theoretically proves that \textit{Choir} \cite{eletreby2017empowering} can only decode 5-10 concurrently transmitting devices. A new distributed coding technique based on CSS is proposed to decode concurrent transmissions below noise floor in a single FFT operation. Experimental results show that this technique can decode 256 concurrent transmissions with 14-62$\times$ improvement in throughput and 15-67$\times$ improvement in latency when compared with existing interference cancellation techniques.  
%\begin{figure}[t]
%	\centering
%	\captionsetup{justification=centering}
%	\includegraphics[width=0.3\textwidth, height=0.3\textwidth, keepaspectratio]{CSMA-CAD}
%	\caption{CSMA-CAD from \cite{liando2019known}}
%	\label{fig:CSMA-CAD}
%\end{figure}

\textbf{Key Insights.} 
The key insights regarding error correction are listed as follows. 
\begin{itemize}
\item An intrinsic property of LoRa to deviate from carrier frequency is identified and exploited in Choir \cite{eletreby2017empowering} to correct errors because of signal collision. Choir is analyzed and its inability to scale-up is identified and a new distributed error correction technique is proposed in Netscatter \cite{hessar2018netscatter} for scalability. 
\item Recently proposed channel coding technique, \textit{DaRe} \cite{marcelis2017dare}, and its analysis \cite{sandell2017application} shows that there is a heavy potential for characterizing efficient channel coding techniques for LoRa to improve error correction, thus improving reception rate and network lifetime.          
\end{itemize}

\subsection{Multiple Access} 
In this subsection, the solutions for multiple access are divided into Link coordination and Resource allocation. The measurements and solutions for each category are explained below with their key insights summarized at the end of each category. 
\subsubsection{Link Coordination}

\textbf{Measurements.} Bankov et. al \cite{bankov2016limits} identifies four important issues. The first issue is whether the gateway should listen to the channel during the interval, T1, between frame reception and transmission of response. There arises a problem if the channel is busy with other scheduled transmissions in that specific interval T1. 
This might cause delay to the ACKs, leading to unwanted retransmissions. 
% Standards do not specify the behaviour of gateways.  
The proposed solution is to cancel the pending transmission that may cause collisions at the end-device and transmit ACKs on the downlink channel. The second issue occurs when two transmitted frames are overlapping in the same time interval over different channels. Gateway would not be able to acknowledge both messages at the specified window with single downlink. The third issue is the limited interval for retransmission. Due to the above said factors, ACKs may take more time to reach an end-device which will increase the retransmission probability even for a successfully delivered message. 
% This limits network scalability. 
The authors recommend increasing the delay interval or use exponential back off to counter this issue. The fourth issue arises when there is no optimal policy to select the data rate for downlink. Simulations conducted show that \textit{packet error rate} and \textit{packet loss ratio} increase with traffic due to improper link coordination.

\textbf{Research Solutions.} Reynders et al. \cite{reynders2018improving} address LoRaWAN's scalability and reliability through a novel MAC protocol, RS-LoRa. This technique works in two phases. 
The gateway sends coarse-grained information of allowed TP and SF for each channel as Class B beacons in the first phase. In the second phase, each end-device selects one parameter combination from the beacon that better suits the node. Assigning different SFs with different  parameter combinations helps to alleviate the Capture Effect. Reliability of network performance is improved by decreasing the Packet Error Ratio up to 20\% than the standard LoRaWAN. This technique proves its superiority to the standard LoRaWAN through NS-3.

Based on measurements obtained through real-world experiments, Haxhibeqiri et al. \cite{haxhibeqiri2017lora} build a simulation model to study the scalability of a single cell LoRaWAN based on interference. It is showed that LoRa physical layer is robust and can send six times traffic than the pure Aloha with 125 KHz bandwidth. Based on end-device density and their data rates given in \cite{huang802proposal}, simulations are carried out to determine node density for different IoT applications.

A series of experiments are conducted to identify the potential of Channel Activity Detection (CAD) and Ideal-CSMA on a dense network of 50 nodes by Liando et al. \cite{liando2019known}. The results show that Ideal-CSMA fails to provide high reception rate when nodes synchronously perform channel detection. Hence, the authors devise a CSMA-CAD with four additional preambles in the packet to achieve doubled PRR. Whenever a preamble is detected, the transmitting SF is randomized and the channel is sensed again for transmission.

The CSMA and CSMA-x are simulated on NS-3. CSMA-x works similar to CSMA but it senses the channel for the gap of \textit{x} ms. The results of this model are compared with the outcomes of real-word test-bed and the results from other works to prove the model's accuracy. It can be seen that even though CSMA cannot provide better results for dense nodes, CSMA-10 is achieving better performance than CSMA. The performance comparison of p-CSMA and CSMA is conducted by Kouvelas et al. \cite{kouvelas2018employing} for small scale networks expound that p-CSMA is an important step to be taken for improving the scalability of LoRa networks but it has not been realized yet on real-world devices.   

There are many retransmission policies devised for wireless networks \cite{levorato2012cognitive, kuo2015qoe, chang2012probabilistic, xiong2008optimal} and wireless sensor networks \cite{naddafzadeh2010distributed, wan2002psfq, wen2007retransmission}. A joint retransmission scheme with compression and channel coding is developed for single-hop networks with energy constraints like LoRa \cite{pielli2017joint}. Its energy efficiency has been theoretically evaluated. This retransmission scheme retransmits the last q failed  data blocks  along with a new data block using compression and coding schemes. Although the performance has been proved theoretically, its implementation with the required computational resource on real-time testbeds will be more helpful to understand the practical efficiency for real-world deployments.

\textbf{Key Insights.} Tailoring different transmission parameter combinations for different end-devices while considering network density is proven to be capable of improving link coordination \cite{reynders2018improving}. Liando et al. \cite{liando2019known} leverage the fact that LoRa gateways can receive packets with different SFs and the same frequency simultaneously to reveal that PRR can be doubled if the SF is randomized on detecting a preamble during CSMA. It is theoretically proven that a retransmission scheme employing data compression and channel coding improves link coordination.         

\subsubsection{Resource Allocation}  
\label{SSS:RA}
\textbf{Measurements.} Semtech's White paper \cite{Semtech2017b} evaluates the capacity of LoRa technology for smart metering applications. This system is deployed in Gehrden, Germany. Around 7000 households were installed with LoRa-enabled smart meters. 11 Kerlink-V1 gateways \cite{Kerlink2016} were mounted on rooftops with 30cm/70cm half wave dipole antenna. A simple meter protocol was employed for reading a seven-digit register. The payload allocates one byte for status and three bytes for each register. The downlink payload length is fixed as 10 bytes. Meters are configured to send unconfirmed payload every 15 minutes and confirmed payload once a day. Class-C end-devices are utilized for this experiment. 24-hour raw data is used for the experiments. It is shown that a gateway with average throughput can handle 470,000 messages per day. It is demonstrated that the network can be scaled up locally by adding gateways. It is also shown that the ADR algorithm improves the network capacity by adjusting the data rate, frame repetition rate and channel allocation. 

 Petajajarvi et al. \cite{petajajarvi2017evaluation} conducted experiments in the University of Oulu, Finland. LoRaMote \cite{Semtech2014} is used as end-device for measurements. These end-devices were configured to send messages to the base station every 5 seconds with no ACKs, no retransmissions and no ADR. End-devices are configured to transmit at six different channels. The gateway is the same as used in \cite{petajajarvi2015coverage}. 
 The packet delivery was above 96.7\% and 95\% when the end-device was static and mobile respectively.
 Similar results have been observed in \cite{Semspace0mmtech} and \cite{magrin2017performance}. Measurements show that most of the campus is covered by SF7 itself. Interesting results were obtained while varying physical parameters. Farthest location is not reached by SF7 and BW125. However, 60\% of the packets were correctly received from the same point with SF7 and BW250. For power consumption evaluations, RN2483 based LoRa module was added to a sensor and actuator kit with a \textit{Keysight's} power analyser. It is noted that energy consumption of the same packet transmission varies by more than 50\% between the maximum and minimum values. This measurement stresses on the usage of ADR to reduce energy consumption.

\indent Angrisani et al. \cite{angrisani2017lora} assess the performance of LoRa under critical noise conditions. Transmitter, powered through a power bank, placed 10m away from the receiver. A White Gaussian noise is generated to corrupt the transmitted signal. The fixed parameters distance, payload and preamble are set to 10m, 1 byte and 8 symbols respectively. On varying SF, BW and CR to all possibilities, authors claim that an increase in BW increases the packet loss with lower SFs. But packet loss is decreased with larger SFs. The authors also state that an increase in CR can trade off an increase in BW and SF. Finally, it is concluded that LoRa is highly robust to high noise levels and recommend further investigation by varying the fixed parameters of this experiment. Iova et al. \cite{iova2017lora} investigate the performance of LoRa in mountain regions and identify the factors affecting transmission parameters. Hakkenberg et al. \cite{hakkenberg2016experimental} and Neumann et al. \cite{neumann2016indoor} evaluate the performance of LoRa in both indoor and outdoor environments and recommend that transmission parameters have to be varied according to the deployed environment. 

\indent Describing the operations of LoRaWAN, Augustin et al. \cite{augustin2016study} evaluate LoRa's receiver sensitivity, network coverage using Freescale KRDM-KL25Z development board \cite{Semiconductors2017} with Semtech 1276 transceiver \cite{Semtech2013} as end-device and Cisco 910 industrial router \cite{Cisco2017} as gateways. Gateway is connected to \textit{The Things Network} server to monitor received packets. Gateway is placed indoors and end-devices were kept moving outdoors in the urban environment. Transmit power of the end-devices was set to minimum 2dBm with a 3-dBi antenna. Packet losses start at 100m. The measured RSSI values were slightly above the specified values for each SF. For network coverage experiments, the gateway was placed 5 m above the ground level and end-devices were kept in a car with default transmit power 14dBm specified in \cite{sornin2015lorawan}. PRR is tested for SF7, SF9, SF12 at various distances with ACK and retransmission turned off. At 2800m, SF7 achieved 0\% PRR while SF12 delivers about 80\% of the packet. The authors find that communication coverage is directly proportional to SF values.

\indent Blenn et al. \cite{blenn2017lorawan} analyse the 9.4GB data obtained from the \textit{The Things Network} during December 2015 and July 2016 from 1618 unique devices. It is inferred that 3.7\% of unique packets were received by two gateways, 1.1\% of unique packets were received by three gateways. Average payload size is 18 Bytes where 93.7\% of captured payloads are less than 50 bytes and 50\% of the payloads are less than 19 bytes. It is observed that using higher SF and higher transmission power results in low packet loss. 

\indent Cattani et al. \cite{cattani2017experimental} conduct experiments to understand the effect of tuning PHY parameters and environmental factors on LoRaWAN communication reliability and energy efficiency. Experimental results show that, for end-devices far away from the gateway, Packet Reception Ratio of fastest PHY setting is only 10\% lower than the slowest setting. Hence, the authors recommend selecting high data rate and high transmission power for the end-devices far away from the gateway. On studying the effect of environmental factors, it is shown that signal strength is decreased by 6dBm at 60\degree C. Even this small deviation can increase packet loss in the messages transmitted by end-devices far away from the gateway. 

\indent Mikhaylov et al. \cite{mikhaylov2017lorawan} study LoRaWAN's susceptibility to inter-network interference. Experiments with and without an interferer between transmitter and receiver gives an insight to design a protocol for finding dynamic communication parameters. Experimental result shows that a personalized communication parameter for each end-device will aid scalability.

\textbf{Research Solutions.} Fair Adaptive Datarate Algorithm (FADR) \cite{abdelfadeel2018fair} is proposed to select SFs and transmission power to achieve data extraction rate among all end-devices. SF is allocated based on the method described in \cite{reynders2017power}, using RSSI and power levels. End-devices are grouped based on regions. This technique simulated in LoRaSim achieves 300\% higher fairness than the technique proposed by Bor et al. \cite{bor2016lora} and 22\% higher fairness than the technique proposed by  Reynders et al. \cite{reynders2017power} while reducing network energy consumption by 22\%.

\indent Bor et al. \cite{bor2016lora} consider bandwidth and transmission power for  scalability analysis through simulation. Georgiou et al. \cite{georgiou2017low} develop a mathematical model for a LoRaWAN network with a single gateway by also considering other unique LoRaWAN features like modulation and radio duty-cycling. A mathematical investigation on link outage, considering signal below SNR threshold and capture effect, is carried out in order to study their effects on scalability. It is inferred that the latter reduces network performance with increase in density of end-devices, which hinders the scalability.

\indent Different from \cite{bor2016lora}, which studies the scalability of LoRaWAN network through LoRaSim, Mikhaylov et al. \cite{mikhaylov2016analysis} present mathematical analysis without considering many factors.  Bor et al. \cite{bor2016lora} build a simulator ruminating Bandwidth and Transmission Power for modeling uplink behavior. Three experiments are conducted. The first experiment with a single gateway and multiple end-devices with homogeneous communication parameter infers that a single gateway can support 120 end-devices per 3.8 hectares. The second experiment contains a single gateway and heterogeneous communication parameters, such that the end-device's uplink air time is decreased, showing 13$\times$ increase in node density than the previous experiment. The final experiment with multiple gateways improves data extraction rate. Two suggested guidelines are to develop a protocol to decide communication parameters dynamically and to evaluate optimal gateway placement for better scalability.

\indent Chall et al. \cite{el2019lorawan} collect empirical data in Lebanon to verify various radio propagation models like Okumura-Hata \cite{hata1980empirical, okumura1968field}, Cost-231 Hata \cite{mogensen1991urban} to find their drawbacks and fix them with additional proposals to make it acceptable for LoRa networks. Bor et al. \cite{bor2016lora} also propose a mathematical model for LoRa  communication coverage based on the empirical data obtained over 2.6 Km of rural area and 100 m of built-up environment. Demetri et al. \cite{demetri2019automated} compare link attenuation of LoRa signals in free space and Bor's model. It is shown that free-space model underestimates signal attenuation while Bor's model overestimates it. It is also claimed that Bor's model \cite{bor2016lora} need on-site measurements which is hard due to that most of the covered regions comprise of transitional links, which are defined as links with dynamic temporal link qualities. Hence, a new automated link quality estimation system without requiring on-site measurements is developed by Demetri et al. \cite{demetri2019automated}. To achieve this, remote sensing spectral images from an open-source satellite is used. These images are fed as input to Support Vector Machine (SVM) \cite{scholkopf2001learning} to classify different constitution of land coverage like water bodies, forests and buildings. Okumura-Hata model \cite{hata1980empirical,okumura1968field} is modified to support LoRa link estimation on different land coverage. The results of SVM classification is used to automatically choose and configure parameters of link estimation model.

\indent Margin et al. \cite{magrin2017performance} implement a new NS-3 module to simulate dense urban environments. The link performance and measurements, signal attenuation due to buildings and other factors are given. 
Spreading Factor assignment is done based on power levels of the end-device at the gateway. 
The gateway will bind with the end-device transmitting on highest received power level. 17 Gateways are placed in a hexagonal grid around the central gateway, covering a 7.5 Km radius. Totally $10^4$ end-devices are placed randomly. The experimental results show that densifying gateways such that each gateway covers 1200m can achieve a packet delivery over 90\%. 
But this increases collision as the number of end-devices using SF7 increases. The authors recommend that ADR mechanisms should be leveraged to counter such collisions. 

\indent Two techniques are proposed by Cumo et. al \cite{cuomo2017explora} to allocate SFs to end-devices. First technique EXPLoRa-SF allocates SFs based on RSSI of the end-device received by the gateway. EXPLoRa-AT guarantees Time-on-Air equalization for all end-devices in addition to SF allocation. This is achieved by \textit{ordered waterfilling} technique to evenly distribute channel load among end-devices in the network. Simulation of EXPLoRa-SF and EXPLoRa-AT in "\textit{LoRaSim}" performs better than the basic ADR.

\indent Bor et al. \cite{bor2017lora} study the impact of transmission parameter selection on communication performance and propose an algorithm to quickly identify the optimal transmission parameters for energy efficiency and reliable communication. It is shown that investing high energy for higher SF values does not always improve communication performance. From experimental results, authors claim that it is also possible to achieve minimum energy efficiency while selecting desired transmission parameters based on application requirements. The proposed probing algorithm finds a transmission parameter that halves the transmission power with PRR larger than a threshold. If not found, other settings that uses at most half the transmission power are probed. If a potential setting is not found, the algorithm employs an iteration bound to try other settings. It is shown that the proposed probing algorithm finds an optimal setting that uses only 44\% more energy than the ideal setting within 285 probes.

\indent Van et al. \cite{van2017scalability} conduct experiment with single, multiple gateways and various SF to study the scalability of LoRaWAN in NS-3 simulator. Error model combined with NS-3 LoRaWAN protocol is constructed through extensive baseband bit error rate simulations and used as an interference model. Experimental results show that usage of ACKs severely affects uplink traffic and having multiple gateways improves scalability to a smaller extent. It is also showed that assigning dynamic communication parameters will help to up-scale node density.

\indent Reynders et al. \cite{reynders2017power} find an optimal SF setting to reduce collision probability and distribute SFs and transmission powers to decrease Packet Error Ratio (PER) of end-devices far away from the gateway. A routine to assign SFs and power control is developed based on genetic algorithm. The key idea of this algorithm is to assign different SFs and power control to different nodes such that signals do not interfere with each other. 
% Four guidelines derived from genetic algorithm ensures the central idea. 
Simulation of this technique in NS-3 shows that the PER of the overall network is reduced by 42\% and the packet error ratio of end-devices far away from the gateway is reduced by 50\%.

\indent Pop et. al \cite{pop2017does} extends \textit{LoRaSim} \cite{bor2016lora} by adding more features like ACK, downlink data messages and presents a new simulator called \textit{LoRaWANSim}. Same experimental settings used in \cite{bor2016lora} are used with additional downlink traffic to study the scalability with downlink ACKs. It is inferred that scalability is hampered as handling ACKs reduce network performance. 

\indent Voigt et al. \cite{voigt2017mitigating} compare the usage of directional antenna and multiple gateways to alleviate interference that arises due to dense deployments. It is shown that the gain of multiple gateways outperforms the usage of directional antennae.

\textbf{Key Insights.} 
The key insights regarding error correction are listed as follows.
\begin{itemize}
\item Communications with higher SF and higher TP can reach a longer distance \cite{blenn2017lorawan, augustin2016study}. 
\item Network density can be scaled-up by adding more gateways and personalizing transmission parameters like data rate and channel allocation using ADR for each end-device \cite{Semtech2017b, mikhaylov2017lorawan, van2017scalability, voigt2017mitigating}. 
\item Energy consumption varies up to 50\% while using lowest and highest transmission power to transmit the same packet. This shows that the lowest possible transmission power should be used for saving maximum energy. 
\item It is not always required to increase SF if some locations cannot reach gateway. Increasing bandwidth on the same SF also increases PRR \cite{petajajarvi2017evaluation}. But this is not always the case at lower SF \cite{angrisani2017lora}. Investing more energy by using high SF does not necessarily improve communication performance \cite{bor2017lora}. Hence, varying transmission parameters based on indoor/outdoor deployed environment gives better performance \cite{hakkenberg2016experimental, neumann2016indoor}. 
\item Only 10\% difference is identified in PRR between fastest and lowest SF and TP settings for the end-devices located at the farthest reachable point from gateways. Hence, lower SF and TP settings can help reduce resource consumption \cite{cattani2017experimental}. 
\item Besides, the communication performance can also be improved by reducing unconfirmed messages \cite{van2017scalability, reynders2017power}.   
\end{itemize}           

\subsection{LoRa Security}

\textbf{Measurements.} On describing the LoRa network stack, Arsas et al. \cite{aras2017exploring} explore the vulnerabilities of LoRa. This paper expounds four possible techniques to compromise the LoRa network. Firstly, compromising security keys. This is easier if an attacker can gain physical access to an end-device. Feasibility of this attack is demonstrated through experiments. Extracting security keys from any end-device will enable the attacker to decrypt any message in the network. Secondly, The Jamming attack. The Jamming attack is an attack in which the communication channel is jammed with an intentional interference to corrupt the data signal sent in that channel. Reynders et al. \cite{reynders2016range} show that LoRa is also prone to Jamming attacks even if the chirp spread spectrum is robust to interferences. Demonstrating this through experiments, it is shown that a specific node can also be targeted. Thirdly, The replay attack. It is an attack in which the intruder intercepts the message and resends it to the receiver whenever the intruder wants to accomplish a particular task. Miller et al. \cite{miller2016lora} show that the consequence of this attack depends on the application scenarios. Finally, the wormhole attack captures a packet from a non-malicious node and this never reaches the server. Some credentials stored in the packet is valid and can be used at any time in the network.      

\indent Butun et al. \cite{butun2018analysis} surveys and verifies the feasible security threats of LoRa V1.1 with Scyther security verification tool \cite{cremers2006scyther}. The attacks verified are (i) RF Jamming attack, (ii) Replay attack, (iii) Class B beacon synchronization attack, (iv) Network traffic analysis, (v) Man-In-The-Middle attack.    

\indent Donmez et al. \cite{donmez2018security} identify the security vulnerabilities of V1.1 in the backward compatibility scenario. The security vulnerabilities of LoRaWAN V1.0.2 and corresponding solutions added in LoRaWAN specification V1.1 to mitigate them are discussed. The open vulnerabilities during backward compatibility scenarios are discussed and countermeasures are proposed to mitigate them. While the specification is explaining only one backward compatibility scenario, this article verifies all possible scenarios to find other vulnerabilities. 

\textbf{Research Solutions.} The discussion on proxy-based key establishment for securing messages in the IoT context in \cite{saied2012hip}, \cite{saied2014lightweight}, \cite{porambage2015proxy}, gives an insight on the proxy-based key establishment for LoRa. Naoui et al. \cite{naoui2016enhancing} discuss the possibilities of applying proxy-based key exchange systems to enhance LoRa security. Bit flipping attack, an attack by which the bits of ciphertext is changed, is countered in \cite{lee2017risk} using circular shift and swap techniques. 

\textit{Key Generation.} Tomasin et al. \cite{tomasin2017security} analyse the security of join procedures, especially through On-The-Air-Activation (OTAA) of LoRa. Join procedures are activated at least once-a-day to check whether the node is still connected to the Network. \textit{DevNonce} is a random number in the \textit{Join-request} message. \textit{DevNonce} should be unique for each \textit{Join Request}. Network Server declines or excludes the node sending \textit{Join-request} with a previously used \textit{DevNonce}. This paper identifies the probability of regenerating a used \textit{DevNonce} and the scenario where malicious node floods Network Server to register possible \textit{DevNonce's} randomly, making the future join request of non-malicious nodes tougher. A random number generator algorithm is proposed and the size of \textit{DevNonce} is increased from 16 to 24 bits to overcome these shortcomings.              

Kim et al. \cite{kim2017simple} focus on resolving the following three problems. The first problem is the current \textit{DevNonce} system. \textit{Join-request} sent by benign end-device can be mistaken as a replay attack by network server because an end-device can regenerate old \textit{DevNonce}. Not storing all the past \textit{DevNonce's} will not prevent replay attack. The second problem is the 24-bit \textit{DevNonce} proposed in the article \cite{tomasin2017security}, that is incompatible with the current LoRaWAN specification. The third problem is the token-based scheme proposed in \cite{na2017scenario}. This prevents replay attack effectively but does not consider the scenario where the token is lost. For example, when the end-device reboots, the token is lost. This paper proposes two types of \textit{Join-requests} called Initial and Non-Initial Join requests. Non-Initial \textit{Join-request} uses the token and it changes after each join procedure is complete. Current \textit{DevNonce} may be regenerated. As the Initial \textit{Join-request} rarely occurs, regeneration of old \textit{DevNonce} is negligible. When the token is lost, the node reboots and initiates Initial \textit{Join-request}. These techniques are proved to enhance security through theory and experiments. 

\indent Oniga et. al \cite{oniga2017analysis} explicates different security aspects of LoRa and proposes a secure network architecture framework. Implementation of this model under different testing scenarios recommend techniques for better data security and privacy of LoRa based applications.       

\textit{Third-party authorisation.} Girard \cite{P.Girard2015} pointed out that both application and network session keys being generated at the network server will create a conflict of interest between network and application service providers. The network server and application server can derive both Application and Network session keys which is not secure if two different organizations are involved. So, a trusted third-party key management architecture is proposed. 

\indent The problem of Key management and update mechanism is well addressed by the techniques proposed in \cite{seo2015effective}, \cite{agrawal2012novel} for wireless sensor networks. As Girard \cite{P.Girard2015} introduces a third-party architecture, communication overhead is increased that may degrade the network performance. So, Kim et al. \cite{kim2017dual} propose a dual key based activation scheme to support key generation and update without adding any complexities. This paper explicates that \textit{AppKey} which is not updated periodically will pose many security threats through which an attacker can steal all transmitted data of a target node. The proposed technique of this paper separates the Network and Application Session key generation to appropriate servers. These keys cannot be derived from a public key and not shared with other devices. This scheme is proven to be both delay and power feasible through experiments.

\textit{Trust mechanism and blockchains.}  Some works \cite{christidis2016blockchains, biswas2016securing, dorri2017blockchain, huh2017managing, samaniego2016using, samaniego2016blockchain}, discuss the application of blockchain technology for IoT scenarios. This is helpful to apply blockchain to LoRa networks. Lin et al. \cite{lin2017using} build a trust mechanism for LoRa using blockchain technology as attacking a blockchain system is computationally difficult as the attacker has to transcend at least half of the system's computational ability. The proposed system implements Blockchain manager component to network server. This framework is proposed for large scale deployments of LoRa like wild-life monitoring, asset tracking and smart parking.   

\textbf{Key Insights.} 
The key insights regarding the security issues are listed as follows.
\begin{itemize}
\item An attacker can decrypt messages in the network by compromising security keys if they can get physical access to end-devices. 
\item LoRa devices are susceptible to jamming attack, replay attack, beacon synchronization attack, traffic analysis and man-in-the-middle attack. 
\item Larger size \textit{DevNonce} prevents join attacks \cite{tomasin2017security}. 
\item Application and Network key generation and update has to be separated based on the application scenario. 
\item Blockchain mechanisms can also be implemented on Low powered devices \cite{lin2017using}.  
\end{itemize}

\section{Open Issues}
\label{SEC:DIRECTION}

As introduced in \ref{SEC:SOLUTION}, various techniques have been proposed to address the challenges of LoRa deployments. Some solutions still leave room to further improve the performance of LoRa. For example, some solutions for choosing dynamic communication parameters consider most of the factors, but did not take into account the ambient temperature which plays a major role in reducing the signal strength \cite {cattani2017experimental}. Based on the above analysis of research challenges and recently proposed solutions in section \ref{SEC:CHALLENGE} and \ref{SEC:SOLUTION}, some open issues of the LoRa technology are presented in this section.    

\subsection{Optimal placement of multiple gateways:} Some works, \cite{Semspace0mmtech}, \cite{bor2016lora}, \cite{van2017scalability}, \cite{voigt2017mitigating} aiding scalability and interference use multiple gateways as a solution. Even though these techniques outperform existing results, using multiple gateways instigate to study the optimal gateway placements for LoRa deployments. Optimal placement of gateways is always dependent on the application and constraints of the hardware used in the application. A generic solution for optimal placement of gateways according to the categories of applications will further improve the performance. 

\subsection{Link Co-ordination:} 

\textbf{Countering degradation due to downlink ACKs:} Some works like \cite{bankov2016limits} and \cite{pop2017does} state that downlink ACKs reduce network performance, as end devices are not able to transmit subsequent packets if downlink ACKs are delayed or corrupted. This gives rise to the need for developing a dynamic ACK mechanism to improve network performance.

\textbf{Dynamic retransmission policies:} One solution to counter downlink ACKs will be setting dynamic retransmission policies. Static retransmission policy degrades the network performance when the time taken by an ACK to reach the end node is larger than the retransmission time. This explains the need for dynamic retransmission policies. A joint retransmission policy with channel coding and compression is theoretically studied, but the implementation of such computationally expensive techniques on constrained LoRa hardware must be considered in future. Besides, it has to be noted that a modular retransmission policy without any dependencies on other techniques is inevitable. None of the techniques has addressed this problem of varying retransmission timers dynamically. The factors triggering retransmission, even in the case of correct reception, must be extensively studied through experiments for various scenarios and a dynamic policy must be devised to improve the performance of LoRa. 

\subsection{Communication Range} Performance evaluations in  \cite{harris2018development,fuidiak2018simulated,petajajarvi2015coverage,petajajarvi2017evaluation,Haxhibeqiri2017,mikhaylov2016analysis} expound the need to provide reliable transmissions for long range LoRa links. Some of the techniques like Du et al. \cite{du2016rateless} improves communication range for wireless sensor networks without heavy hardware modifications. The communication range could be further enlarged in the future as LoRa chips will evolve to support new functionalities.

\indent Choir \cite{eletreby2017empowering} improves the range but its implementation on commercial radio chips is ambiguous as they were experimented on USRP radios and may require modifying the commercial radio. NetScatter \cite{hessar2018netscatter} theoretically proves that Choir \cite{eletreby2017empowering} cannot scale-up well. NetScatter \cite{hessar2018netscatter} leverages backscatter with distributed coding to enable concurrent transmission of 256 nodes which still needs some hardware modifications on commercial chips. Even though the above techniques Choir \cite{eletreby2017empowering} and NetScatter \cite{hessar2018netscatter} have improved the range, communication range still needs to be improved \textit{without heavy modifications of the commercially available chips}.    

\subsection{Security} Security spans over a range of attacks like node impersonation, eavesdropping, Black hole attack, Wormhole attack, etc., as discussed by Zhou et al. \cite{zhou2008securing}. Only few key management techniques have been discussed and proposed for LoRa. Several attacks still need to be addressed to secure LoRa networks. Even though some techniques enhance the security of existing LoRa standard, security requirements of LoRa are not discussed based on the applications of LoRa. In the future, each application of LoRa will demand their own security needs. For example, some applications may require Network and Application Session keys to be independent and Network Session Key should be confidential from application server and vice-versa. Hence, the security needs of each deploying scenario have to be deeply investigated to mitigate vulnerabilities arising due to different application scenarios. 
\section{Conclusion}
\label{SEC:CONCLUSION}

Among different LPWAN technologies, LoRa networking is widely adopted, since it allows to build and maintain an autonomous network without third-party infrastructure, while satisfying the low power and long range communication requirements. By investigating the challenges faced during deploying LoRa networks, recent solutions developed are discussed in detail. Based on the challenges and recent solutions, we present some open issues that need to be addressed for practical large-scale deployment of LoRa networks.      
\balance
\bibliographystyle{IEEEtran}
\bibliography{ref}

\end{document}